\documentclass{ws-mpla}
\usepackage[super]{cite}
\usepackage{graphicx}

\usepackage{latexsym}
\usepackage{amsfonts}
\usepackage{amsmath}
\usepackage{amssymb}
\usepackage{revsymb}
\usepackage{bm}
\usepackage{subfigure}
\usepackage[citecolor=green]{hyperref} 

\makeatletter
\newcommand{\xRightarrow}[2][]{\ext@arrow 0359\Rightarrowfill@{#1}{#2}}
\makeatother

\begin{document}
\markboth{Aydemir, Minic, Sun, Takeuchi}
{750 GeV diphoton excess in unified $SU(2)_L\times SU(2)_R\times SU(4)$ models from NCG}

\catchline{}{}{}{}{}

\title{The 750 GeV diphoton excess in unified $SU(2)_L\times SU(2)_R\times SU(4)$ models from noncommutative geometry}

\author{Ufuk Aydemir}
\address{Department of Physics and Astronomy, Uppsala University, 
SE-751 20 Uppsala, Sweden\\
ufuk.aydemir@physics.uu.se}

\author{Djordje Minic, Chen Sun, and Tatsu Takeuchi}
\address{Center for Neutrino Physics, Department of Physics, Virginia Tech, Blacksburg, VA 24061 USA\\
dminic@vt.edu, sunchen@vt.edu, takeuchi@vt.edu}

\maketitle

\pub{Received (Day Month Year)}{Revised (Day Month Year)}

\begin{abstract}
We discuss a possible interpretation of the $750$ GeV diphoton resonance, recently reported at the LHC, 
within a class of $SU(2)_L\times SU(2)_R\times SU(4)$ models with gauge coupling unification.
The unification is imposed by the underlying non-commutative geometry (NCG), which in these models is
extended to a left-right symmetric completion of the Standard Model (SM).
Within such \textit{unified} $SU(2)_L\times SU(2)_R\times SU(4)$ models the Higgs content is restrictively determined
from the underlying NCG, instead of being arbitrarily selected.
We show that the observed cross sections involving the $750$ GeV diphoton resonance could be realized through a SM singlet scalar field accompanied by colored scalars, present in these unified models. 
In view of this result we discuss the underlying rigidity of these models in the NCG framework and the wider implications of the NCG approach for physics beyond the SM.

\end{abstract}
 
\ccode{PACS Nos.: 12.10.Dm,12.60.-i,12.90.+b}

\section{Introduction}

After the discovery of the Higgs boson, the obvious burning question of particle physics is: what lies beyond the completed physics of the Standard Model (SM)?
For many years, theoretical extensions of the SM have been driven by the hard lessons learned
(and perhaps prejudices acquired) 
in the context of effective field theory, the underlying language of the SM.
In these discussions, which include supersymmetry, technicolor, and extra dimensions, among many others, 
the theoretical motivations were rooted in deep questions such as the hierarchy problem, or the idea of grand unification (GUT).\footnote{%
For a review see, for example, Ref.~\refcite{Mohapatra:2002book}.
}
More recently, growing attention has been devoted to the underlying non-commutative geometry (NCG) of the Standard Model,\footnote{%
For recent reviews see Refs.~\refcite{Chamseddine:2010ud,Chamseddine:2012sw}.} 
with implications for its natural left-right symmetric completion,
as discussed in Refs.~\refcite{Chamseddine:2013rta} and \refcite{Chamseddine:2015ata}.
In the NCG framework, the SM itself harbors a GUT-like structure, and its natural completion appears to be 
a GUT-like \textit{unified} $G_{224}=SU(2)_L\times SU(2)_R\times SU(4)$ model. 
The question is: are there any hints for this unified 
$G_{224}$ NCG framework
in the observed world, as opposed to the canonical non-restricted
versions?\cite{Mohapatra:2002book,Pati:1974yy,Mohapatra:1974hk,Mohapatra:1974gc,Senjanovic:1975rk,Mohapatra:1980qe}\footnote{%
See Refs.~\refcite{Aydemir:2015nfa,Aydemir:2013zua,Aydemir:2014ama} for recent discussions. 
}

Recently, ATLAS\cite{ATLAS:2015diphoton} and CMS\cite{CMS:2015dxe}
have both reported a resonance in the diphoton channel with an invariant mass around $750$ GeV.
The local significances were respectively $3.6\sigma$ and $2.6\sigma$,
assuming a narrow width resonance.
These signals may be the first hint associated with the long-anticipated physics beyond the SM.
The 95\% CL cross section upper limit around 750 GeV set by ATLAS (CMS) is roughly $10\pm 2.8$ fb ($6.5\pm 3.5$ fb) using 3.2 fb$^{-1}$ (2.6 fb$^{-1}$) of data at 
$\sqrt{s}=13\,\mathrm{TeV}$, assuming the resonance is a scalar produced through gluon-gluon fusion. 
When the width of the resonance is allowed to vary,
a maximum local significance of $3.9\sigma$ is attained by ATLAS at a width of 45 GeV. 
On the other hand, the local significance attains its maximum for
a narrow width resonance in the CMS results. Therefore, at this stage, given these preliminary analyses, it is difficult to infer conclusively whether the width of the resonance is wide or narrow\cite{Buckley:2016mbr}.

\begin{figure}[!t]
\begin{center}
\includegraphics[scale=0.9]{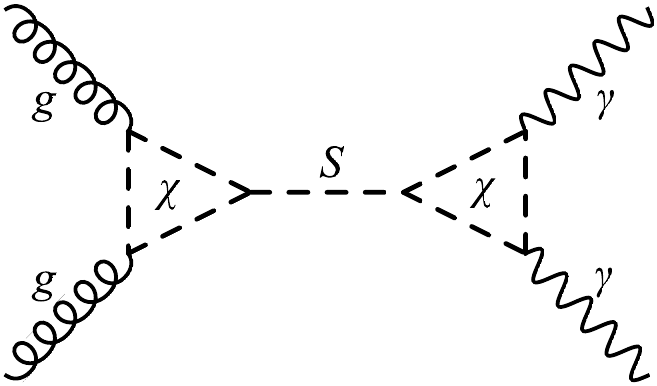}
\end{center}
\caption{The Feynman diagram of the production and decay of the SM-singlet scalar 
$\mathcal{S}$ at the LHC through colored-scalar $\chi$ in the loop.}
\label{fig:PD}
\end{figure}

In this letter, we discuss a possible identification of this resonance with SM-singlet scalars
in the NCG motivated unified $G_{224}$ models.\cite{Chamseddine:2013rta,Chamseddine:2015ata}
A plausible and economical way to realize the LHC diphoton signal in the unified $G_{224}$ context 
is to couple this SM-singlet scalar to gluons and photons via loops of colored scalars, as recently discussed in Ref.~\refcite{Aydemir:2016qqj} in the context of $SO(10)$ GUT, cf. Fig.~\ref{fig:PD}.\footnote{%
Dasgupta et al. in Ref.~\refcite{Dasgupta:2015pbr} have shown
that coupling the SM-singlet scalar to quarks and photons via mixing with
the SM Higgs boson would lead to too small a cross section.
}
The NCG models we consider have either an $SU(2)_R$ triplet
$\Delta_R(1,3,10)$, or an $SU(2)_R$ doublet $\widetilde{\Delta}_{R}(1,2,4)$
in their scalar content, where the three numbers refer to the
dimensions of the $G_{224}=SU(2)_L\times SU(2)_R\times SU(4)_C$ irreducible representations.
The SM-singlet scalar $\mathcal{S}$, which we identify with the $750$~GeV resonance, is assumed to be the excitation of the electrically neutral component 
$\Delta_{R1}^{0}$ of $\Delta_R(1,3,10)$, 
or that of $\widetilde{\Delta}_R(1,2,4)$ (denoted as $\widetilde{\Delta}_{R1}^{0}$), depending on the model considered. 
$\Delta_{R1}^{0}$ (or $\widetilde{\Delta}_{R1}^{0}$) is also the field that breaks the gauge symmetry of the $G_{224}$ model to that of the SM, by acquiring a vacuum expectation value (VEV) at the scale $M_C$ where
$G_{224}$ breaks to $G_{213}=SU(2)_L\times U(1)_Y\times SU(3)_C$ of the SM. 
A color-triplet components of $\Delta_R(1,3,10)$ 
(or $\widetilde{\Delta}_R(1,2,4)$) is assumed to
survive down to low energies (TeV-scale) to take on the role of the
$\chi$-field in Fig.~\ref{fig:PD}.
This is but one way that one could embed the 750 GeV diphoton resonance into the
NCG $G_{224}$ framework, and we use this as a demonstrative example.

While this identification itself is fairly straightforward, 
and it can already be inferred from the similar $SO(10)$ analysis of Ref.~\refcite{Aydemir:2016qqj} 
that the cross section and width can be made to come out in the right ballpark,
the question is whether the assumed 
symmetry breaking and scalar survival scenario can actually be realized in the NCG $G_{224}$ models,
given the gauge-coupling-unification requirement and restricted
scalar content which limits our ability to adjust the renormalization group running of those couplings.
Indeed, we have demonstrated in Ref.~\refcite{Aydemir:2015nfa} that
realizing a 2~TeV $W_R$, which had been suggested by the LHC data\cite{Aad:2014aqa,Aad:2015owa,Aad:2015ipg,Khachatryan:2014gha,Khachatryan:2015sja,Khachatryan:2016yji}, 
in the same NCG models was highly non-trivial due to the unification requirement
applying conflicting pressure on the symmetry breaking scales.
Thus, we subject our scenarios to renormalization group equation (RGE) analyses
to check their feasibilities.

The main message of this letter is that even though the $750$ GeV diphoton resonance can be accommodated within the NCG motivated unified $G_{224}$ models, the price one has to pay is
a certain amount of fine tuning in the sector involving the necessary colored scalars. 
This is somewhat similar to the main message of Ref.~\refcite{Aydemir:2015nfa}, and points to the underlying rigidity of the NCG framework.
However, this conclusion is based on effective-field-theory reasoning, 
which could fail in the NCG framework due to the possible mixing between the short-distance and long-distance physics as discussed in our previous work.\cite{Aydemir:2013zua,Aydemir:2014ama,Aydemir:2015nfa}

This letter is organized as follows. 
In section~2, we present the list of the NCG based unified $G_{224}$ models that are analyzed, and discuss how the $750$ GeV diphoton resonance could be explained within their
framework.
In section~3, we address the question of whether the unification of gauge couplings 
can be achieved naturally in those models.
We conclude in section~4 with an outlook on the rigid phenomenological aspects of the NCG framework. 
In the process, we follow the technology discussed in our previous paper\cite{Aydemir:2015nfa} to which we refer the reader for further technical details.

\section{Diphoton resonance in NCG based unified $G_{224}$ models}

\begin{table}[ht!]
\tbl{The scalar content of the three NCG based unified $G_{224}$ models
proposed by Chamseddine, Connes, and van Suijlekom in Refs.~4 and 5,
compared to the scalar content of the $SO(10)$ based $G_{224}$ model, 
discussed in Ref.~16, below its unification scale where 
the $SO(10)$ symmetry is broken to $G_{224}$.  
}
{\begin{tabular}{l|l|l}
\hline
Model & Symmetry & Higgs Content \vphantom{\Big|}\\
\hline\hline
A & $G_{224}$  & $\phi(2,2,1)$, $\widetilde{\Delta}_R(1,2,4)$, $\Sigma(1,1,15)$ \vphantom{\bigg|}\\
\hline
B & $G_{224}$  & $\phi(2,2,1)$, $H(1,1,6)$, $\Delta_R(1,3,10)$, $\widetilde{\Sigma}(2,2,15)$ \vphantom{\bigg|}\\
\hline
C & $G_{224D}$ & $\phi(2,2,1)$, $H(1,1,6)\times 2$, $\Delta_R(1,3,10)$, $\Delta_L(3,1,10)$, $\widetilde{\Sigma}(2,2,15)$ \vphantom{\bigg|}\\
\hline\hline
$SO(10)$ & $G_{224}$ &  $\phi(2,2,1)$, $\Delta_R(1,3,10)$, $\Sigma(1,1,15)$ \vphantom{\bigg|}\\
\hline
\end{tabular}}
\label{NCG-HiggsContent}
\end{table}

In this section, we list the three unified $G_{224}$ models proposed by
Chamseddine, Connes, and van Suijlekom in Refs.~\refcite{Chamseddine:2013rta} and \refcite{Chamseddine:2015ata}, and specify how we fit the diphoton resonance into
their particle content. 
These models emerge from an underlying NCG, which is an extension of the NCG of the SM to that of
left-right symmetric models.
The three versions differ in the scalar sector content,
and the unbroken symmetry structure as listed in Table~\ref{NCG-HiggsContent}.
We use the following notation for the symmetries:
\begin{eqnarray}
G_{224D} & = & SU(2)_L\otimes SU(2)_R\otimes SU(4)_C\otimes D\;,\vphantom{\Big|}\cr
G_{224}  & = & SU(2)_L\otimes SU(2)_R\otimes SU(4)_C\;,\vphantom{\Big|}\cr
G_{213}  & = & SU(2)_L\otimes U(1)_{Y}\otimes SU(3)_C\;,\vphantom{\Big|}\cr
G_{13}   & = & U(1)_{Q}\otimes SU(3)_C\;,\vphantom{\Big|}
\label{Gdef}
\end{eqnarray}
where $D$ in $G_{224D}$ refers to the left-right symmetry, a $Z_2$ symmetry which keeps the left and the right sectors equivalent. The last row of Table~\ref{NCG-HiggsContent} lists the scalar content of
an $SO(10)$ based $G_{224}$ model studied in Ref.~\refcite{Aydemir:2016qqj},
below its unification scale where the $SO(10)$ had broken to $G_{224}$.  
The scalars $\phi(2,2,1)$, $\Delta_R(1,3,10)$, $\Sigma(1,1,15)$ are respectively obtained from the $SO(10)$ multiplets $\mathbf{10}$ (or $\mathbf{120}$), $\mathbf{126}$, and $\mathbf{210}$.
The $\mathbf{210}$ also includes a $(1,1,1)$ representation, whose VEV breaks
$SO(10)$ down to $G_{224}$.\cite{Slansky:1981yr}

As in our previous paper \cite{Aydemir:2015nfa} we will not attempt to review the NCG foundations of these models or to justify their derivation, 
but simply look at their consequences from a purely phenomenological viewpoint in 
the light of the possible $750$ GeV diphoton resonance. 
The distinguishing feature of NCG motivated versions of the SM\cite{Chamseddine:2010ud,Chamseddine:2012sw} as well as its 
$G_{224}$ completion discussed here is that they come with GUT-like
coupling unification conditions,
due to the underlying spectral action having only one overall coupling. 
This is not the case for the canonical $G_{224}$ constructions found in the literature.\cite{Mohapatra:2002book,Pati:1974yy,Mohapatra:1974gc,Mohapatra:1974hk,Senjanovic:1975rk}

The decompositions of the various scalar fields, which appear in Table~\ref{NCG-HiggsContent},
into irreducible representations of the subgroups as the symmetry breaks
from $G_{224}$ (or $G_{224D}$) to $G_{2213}$ and then to $G_{213}$ are shown in
Table~\ref{Decompositions}.
In model A, we assume that $G_{224}$ is broken directly to $G_{213}$ by
$\widetilde{\Delta}_{R1}^0(1,0,1)$ acquiring a VEV, and
$\mathcal{S}$ is identified with the excitation of $\widetilde{\Delta}_{R1}^0$.
In models B and C, we assume that $G_{224}/G_{224D}$ is broken directly to
$G_{213}$ by $\Delta_{R1}^0(1,0,1)$ acquiring a VEV, while
$\mathcal{S}$ is identified with the excitation of $\Delta_{R1}^0$.
In all three models, the colored field $\Delta_{R3}^{-2/3}(1,-4/3,3)$,
which is contained in the decompositions of both $\widetilde{\Delta}_{R}(1,2,4)$
and $\Delta_{R}(1,3,10)$, is assumed to survive 
below the symmetry breaking scale.

\begin{table}[ht!]
\tbl{The decomposition of various $G_{224}$ representations into those of $G_{2213}$ and $G_{213}$ (SM).}
{\begin{tabular}{l|l|l}
\hline
$G_{224}$ & $G_{2213}$ & $G_{213}$ \\
\hline\hline
$\phi(2,2,1)$ 
& $\phi(2,2,0,1)$ 
& $\phi_2(2,1,1)$, $\phi_2'(2,-1,1)$ \vphantom{\Bigg|}\\
\hline
$\widetilde{\Delta}_R(1,2,4)$
& $\widetilde{\Delta}_{R1}(1,2,1,1)$ 
& $\widetilde{\Delta}_{R1}^{0}(1,0,1)$, $\widetilde{\Delta}_{R1}^{+}(1,2,1)$ \vphantom{\Bigg|}\\
\cline{2-3}
& $\widetilde{\Delta}_{R3}\left(1,2,-\dfrac{1}{3},3\right)$ 
& $\widetilde{\Delta}_{R3}^{1/3}\left(1,\dfrac{2}{3},3\right)$, 
  $\widetilde{\Delta}_{R3}^{-2/3}\left(1,-\dfrac{4}{3},3\right)$ \vphantom{\Bigg|}\\
\hline
$\Delta_R(1,3,10)$
& $\Delta_{R1}(1,3,2,1)$
& $\Delta_{R1}^{0}(1,0,1)$, $\Delta_{R1}^{+}(1,2,1)$, $\Delta_{R1}^{++}(1,4,1)$ \vphantom{\Bigg|}\\
\cline{2-3}
& $\Delta_{R3}\left(1,3,\dfrac{2}{3},3\right)$
& $\Delta_{R3}^{+4/3}\left(1,\dfrac{8}{3},3\right)$,
  $\Delta_{R3}^{+1/3}\left(1,\dfrac{2}{3},3\right)$,
  $\Delta_{R3}^{-2/3}\left(1,-\dfrac{4}{3},3\right)$ \vphantom{\Bigg|}\\
\cline{2-3}
& $\Delta_{R6}\left(1,3,-\dfrac{2}{3},6\right)$
& $\Delta_{R6}^{+2/3}\left(1,\dfrac{4}{3},6\right)$,
  $\Delta_{R6}^{-1/3}\left(1,-\dfrac{2}{3},6\right)$,
  $\Delta_{R6}^{-4/3}\left(1,-\dfrac{8}{3},6\right)$ \vphantom{\Bigg|}\\
\hline
$\Delta_L(3,1,10)$
& $\Delta_{L1}(3,1,2,1)$
& $\Delta_{L1}(3,2,1)$ \vphantom{\Bigg|}\\
\cline{2-3}
& $\Delta_{L3}\left(3,1,\dfrac{2}{3},3\right)$
& $\Delta_{L3}\left(3,\dfrac{2}{3},3\right)$ \vphantom{\Bigg|}\\
\cline{2-3}
& $\Delta_{L6}\left(3,1,-\dfrac{2}{3},6\right)$
& $\Delta_{L6}\left(3,-\dfrac{2}{3},6\right)$ \vphantom{\Bigg|}\\
\hline
$H(1,1,6)$
& $H_3\left(1,1,\dfrac{2}{3},3\right)$
& $H_3^{1/3}\left(1,\dfrac{2}{3},3\right)$ \vphantom{\Bigg|} \\
\cline{2-3}
& $H_{\bar{3}}\left(1,1,-\dfrac{2}{3},\bar{3}\right)$
& $H_{\bar{3}}^{-1/3}\left(1,-\dfrac{2}{3},\bar{3}\right)$ \vphantom{\Bigg|} \\
\hline
$\Sigma(1,1,15)$ 
& $\Sigma_1(1,1,0,1)$ 
& $\Sigma_1^0(1,0,1)$ \vphantom{\Bigg|}\\
\cline{2-3}
& $\Sigma_3\left(1,1,-\dfrac{4}{3},3\right)$ 
& $\Sigma_3^{-2/3}\left(1,-\dfrac{4}{3},3\right)$ \vphantom{\Bigg|}\\ 
\cline{2-3}
& $\Sigma_{\bar{3}}\left(1,1,\dfrac{4}{3},\bar{3}\right)$
& $\Sigma_{\bar{3}}^{2/3}\left(1,\dfrac{4}{3},\bar{3}\right)$ \vphantom{\Bigg|}\\
\cline{2-3}
& $\Sigma_8(1,1,0,8)$ 
& $\Sigma_8^0(1,0,8)$ \vphantom{\Bigg|}\\ 
\hline
$\widetilde{\Sigma}(2,2,15)$
& $\widetilde{\Sigma}_1(2,2,0,1)$
& $\widetilde{\Sigma}_1(2,1,1)$, $\widetilde{\Sigma}_1'(2,-1,1)$ \vphantom{\Bigg|}\\
\cline{2-3}
& $\widetilde{\Sigma}_3\left(2,2,-\dfrac{4}{3},3\right)$ 
& $\widetilde{\Sigma}_3\left(2,-\dfrac{7}{3},3\right)$, $\widetilde{\Sigma}_3'\left(2,-\dfrac{1}{3},3\right)$ \vphantom{\Bigg|}\\
\cline{2-3}
& $\widetilde{\Sigma}_{\bar{3}}\left(2,2,\dfrac{4}{3},\bar{3}\right)$ 
& $\widetilde{\Sigma}_{\bar{3}}\left(2,\dfrac{7}{3},\bar{3}\right)$, $\widetilde{\Sigma}_{\bar{3}}'\left(2,\dfrac{1}{3},\bar{3}\right)$ \vphantom{\Bigg|}\\
\cline{2-3}
& $\widetilde{\Sigma}_8(2,2,0,8)$ 
& $\widetilde{\Sigma}_8(2,1,8)$, $\widetilde{\Sigma}_8'(2,-1,8)$ \vphantom{\Bigg|}\\ 
\hline
\end{tabular}}
\label{Decompositions}
\end{table}

The advantage of this choice of the surviving colored scalar is 
that it exists in all three models, and that it is similar to the one considered in the $SO(10)$ context in Ref.~\refcite{Aydemir:2016qqj}, where the reproducibility of the recently reported LHC signal
has been demonstrated with such a new degree of freedom. The other colored components of $\widetilde{\Delta}_{R}(1,2,4)$
and $\Delta_{R}(1,3,10)$ could also serve this end.
Single step breaking from $G_{224}$ to $G_{213}$ is assumed for the sake of simplicity,\footnote{%
In models B and C, the breaking sequence $G_{224}\rightarrow G_{2213}\rightarrow G_{213}$ 
considered in Ref.~\refcite{Aydemir:2015nfa} requires scalar composites acquiring a VEV in the intermediate steps.}
and also due to our experience in Ref.~\refcite{Aydemir:2015nfa} 
telling us that introducing multi-step breaking does not necessarily facilitate the grafting
of the NCG models to the SM at low energies.

In the $SO(10)$ model of Ref.~\refcite{Aydemir:2016qqj}, the $750\,\mathrm{GeV}$ resonance $\mathcal{S}$ was identified with the excitation of the charge neutral component $\Delta_{R3}^0$ of $\Delta_R(1,3,10)$, which acquires a VEV breaking $G_{2213}$ down to $G_{213}$ at $M_R = 5\,\mathrm{TeV}$, and
only one of the colored components, $\chi = \Delta_{R3}^{-2/3}(1,-4/3,3)$, was assumed to survive
below this breaking.
This is the exact same identification as in models B and C, except the assumed symmetry breaking
pattern is different.
Since $\mathcal{S}$ is a singlet under the SM gauge group $G_{213}$, 
it cannot directly couple to gluons or photons. 
The coupling is induced by $\chi$-loops as shown in Fig.~\ref{fig:PD}.
Assuming a coupling between $\mathcal{S}$ and $\chi$ of the form
\begin{equation}
\kappa M_R\mathcal{S}\chi^\dagger\chi\;,\qquad M_R\,=\,5\,\mathrm{TeV}\;,
\end{equation}
where $\kappa$ is a dimensionless parameter, and $M_\chi > M_\mathcal{S}/2$ so that $\mathcal{S}$
does not decay into a $\chi$ pair, it has been shown in Ref.~\refcite{Aydemir:2016qqj} that the LHC signal can be reproduced for a wide range of $(\kappa,M_\chi)$ values around
$\kappa=O(1)$ and $M_\chi=O(1\,\mathrm{TeV})$.
Thus, without repeating the analysis we conclude that our NCG models
can also reproduce the LHC signal provided a similar coupling exists between $\mathcal{S}$ and $\chi$, and the assumed particle content allows the required unification
of gauge couplings at a high scale.

Several comments are in order.
The $\Delta_R(1,3,10)$ scalar
is associated with a rich phenomenology as discussed by Mohapatra and Marshak in Ref.~\refcite{Mohapatra:1980qe}, including the generation of Majorana neutrino mass and
neutron-anti-neutron oscillations.
These depend on the Yukawa couplings of the $\Delta_R(1,3,10)$ to the fermions, and
the quartic coupling of the $\Delta_R(1,3,10)$ to itself.
In the NCG approach, the Dirac operator, which includes the Yukawa couplings,
is the input from which the entire model is constructed.
The scalar content of the model as well as their quartic couplings are derived
from the Dirac operator.\footnote{%
See appendix of Ref.~\refcite{Chamseddine:2013rta}.
}
Therefore, the NCG approach can, in principle, make predictions in regards to
neutron-anti-neutron oscillations.
However, it is necessary to check the viability of the model before preforming
such detailed analyses, so this will not be discussed further in this paper.

\section{Unification of the couplings}

As discussed in the introduction, the unification of couplings in the NCG based
$G_{224}$ models imposes non-trivial requirements on the symmetry breaking scales,
given that the scalar content of each model is restricted and cannot be
changed at will. 
In this section, we discuss whether the unification of the couplings can be achieved
in the NCG based $G_{224}$ models with the assumed particle content and 
scalar survival assumptions.
In contrast to our work in Ref.~\refcite{Aydemir:2015nfa}, 
we assume direct breaking of $G_{224}$ to $G_{213}$ at a single scale $M_C$, 
between the unification scale $M_U$ and the 
electroweak symmetry breaking (EWSB) scale $M_Z$.
Between the scales $M_C$ and $M_Z$, in addition to the usual SM particle content
we have the $\chi=\Delta_{R3}^{-2/3}\left(1,-4/3,3\right)$ field contributing to the RGE,
which we assume is the only colored scalar to survive below $M_C$, and possess a mass of around a TeV.  The $750\,\mathrm{GeV}$ scalar $\mathcal{S}$ is an SM singlet and consequently
does not contribute to the RG running of the SM gauge couplings.

\subsection{Boundary and Matching Conditions} 

The symmetry breaking chain of the model considered in this letter 
has been discussed in detail in our previous papers\cite{Aydemir:2015nfa, Aydemir:2015oob}. The ordering of the breaking scales must be strictly maintained in the computations, \textit{i.e.}
\begin{equation}
M_U \ge\; M_C  \ge\; M_Z\ .
\end{equation}
We label the energy intervals in between symmetry breaking scales
$[M_Z,M_C]$ and $[M_C,M_U]$ with Roman numerals as
%
\begin{eqnarray}
\mathrm{I}  & \;:\; & [M_Z,\;M_C]\;,\quad G_{213}  \;(\mathrm{SM}) \;,\cr
\mathrm{II} & \;:\; & [M_C,\;M_U]\;,\quad G_{224}\quad\mbox{or}\quad G_{224D} \;.
\label{IntervalNumber}
\end{eqnarray}
The boundary/matching conditions we impose on the couplings at the symmetry breaking scales are:
\begin{eqnarray}
M_U & \;:\; & g_L(M_U) \;=\; g_R(M_U) \;=\; g_4(M_U) \;, \vphantom{\bigg|} 
\cr
M_C & \;:\; & \sqrt{\frac{2}{3}}\,g_{BL}(M_C) \;=\; g_3(M_C)=g_4(M_C) \;,\quad g_2(M_C)\;=\;g_L(M_C)\;,\cr
 & \;\; & \frac{1}{g_1^2(M_C)} \;=\; \frac{1}{g_R^2(M_C)}+ \frac{2}{3}\frac{1}{g_4^2(M_C)}\;,\quad\;
\vphantom{\Bigg|}
\cr
M_Z & \;:\; & \frac{1}{e^2(M_Z)} \;=\; \frac{1}{g_1^2(M_Z)}+\frac{1}{g_2^2(M_Z)}\;.\vphantom{\Bigg|} 
\label{Matching}
\end{eqnarray}
The low energy data which we will use as boundary conditions to the RG running are
\cite{Agashe:2014kda,ALEPH:2005ab}
\begin{eqnarray}
\alpha(M_Z) \;=\; 1/127.9\;,\quad
\alpha_s(M_Z) \;=\; 0.118\;,\quad
\sin^2\theta_W(M_Z) \;=\; 0.2312\;,
\label{SMboundary}
\end{eqnarray}
at $M_Z=91.1876\,\mathrm{GeV}$, which translates to
\begin{equation}
g_1(M_Z) \;=\; 0.36\;,\quad
g_2(M_Z) \;=\; 0.65\;,\quad
g_3(M_Z) \;=\; 1.22\;.
\label{MZboundary}
\end{equation}
Note that the coupling constants are all required to remain in the perturbative regime during the
evolution from $M_U$ down to $M_Z$.

\subsection {One-loop renormalization group running}

For a given particle content; the gauge couplings, in an energy interval $\left[M_A,M_B\right]$, are evolved  according to the 1-loop RG relation
\begin{eqnarray}
\frac{1}{g_{i}^{2}(M_A)} - \dfrac{1}{g_{i}^2(M_B)}
\;=\; \dfrac{a_i}{8 \pi^2}\ln\dfrac{M_B}{M_A}
\;,
\end{eqnarray}
where the RG coefficients $a_i$ are given by \cite{Jones:1981we,Lindner:1996tf}
\begin{eqnarray}
\label{1loopgeneral}
a_{i}
\;=\; -\frac{11}{3}C_{2}(G_i)
& + & \frac{2}{3}\sum_{R_f} T_i(R_f)\cdot d_1(R_f)\cdots d_n(R_f) \cr
& + & \frac{\eta}{3}\sum_{R_s} T_i(R_s)\cdot d_1(R_s)\cdots d_n(R_s)\;.
\end{eqnarray}
The summation in Eq.~(\ref{1loopgeneral}) is over irreducible chiral representations of fermions ($R_f$) in the second term and those of scalars ($R_s$) in the third. The coefficient $\eta$ is either 1 or 1/2, depending on whether the corresponding representation is complex or real, respectively. $C_2(G_i)$ is the quadratic Casimir for the adjoint representation of the group $G_i$,
and $T_i$ is the Dynkin index of each representation. For $U(1)$ group, $C_2(G)=0$ and
\begin{equation}
\sum_{f,s}T \;=\; \sum_{f,s}\left(\dfrac{Y}{2}\right)^2\;,
\label{U1Dynkin}
\end{equation}
where $Y/2$ is the $U(1)$ charge, the factor of $1/2$ coming from the traditional
normalizations of the hypercharge and $B-L$ charges.

The RG coefficients, $a_i$, differ depending on the particle content in each energy interval, changing every time symmetry breaking occurs. We will distinguish the $a_i$'s in different intervals with the corresponding roman numeral superscript, cf. Eq.~(\ref{IntervalNumber}).
Together with the matching and boundary conditions of Eqs.~(\ref{Matching}), (\ref{SMboundary}),
(\ref{MZboundary}), 1-loop RG running leads to the following
conditions on the symmetry breaking scales $M_U$ and $M_C$:
\begin{eqnarray}
2\pi\left[\dfrac{3-8\sin^2\theta_W(M_Z)}{\alpha(M_Z)}\right]
& = & 
 \left(3a_1 -5a_2\right)^\mathrm{I}\ln\dfrac{M_C}{M_Z}
+\left(-5a_L+3a_R+2a_4\right)^\mathrm{II}\ln\dfrac{M_U}{M_C}
\;,
\vphantom{\Bigg|}
\cr
2\pi\left[\dfrac{3}{\alpha(M_Z)} - \dfrac{8}{\alpha_s(M_Z)}\right]
& = &
 \left(3a_1 + 3a_2 - 8a_3\right)^\mathrm{I}\ln\dfrac{M_C}{M_Z}
+\left(3a_L+3a_R-6a_4\right)^\mathrm{II}\ln\dfrac{M_U}{M_C}
\;.
\vphantom{\Bigg|}
\cr
& &
\end{eqnarray}
%
The unified coupling $\alpha_U$ at scale $M_U$ can then be obtained from
%
\begin{eqnarray}
\label{A6}
\dfrac{2\pi}{\alpha_U}
 =  \dfrac{2\pi}{\alpha_s(M_Z)}
-\left( a_4^\mathrm{II}\;\ln\dfrac{M_U}{M_C}
+ a_3^\mathrm{I}\;\ln\dfrac{M_C}{M_Z}
\right)
\;.
\end{eqnarray}
Thus, once the RG coefficients in each interval are specified,
the scales $M_U$ and $M_C$, and the value of $\alpha_U$ will be uniquely determined.
For the computations to be meaningful, however, $M_U$ must stay below the Planck scale, 
and $\alpha_U$ must be in the perturbative regime.

\subsection{Results}

\begin{figure}[ht!]
\subfigure[ ]{\includegraphics[width=6cm]{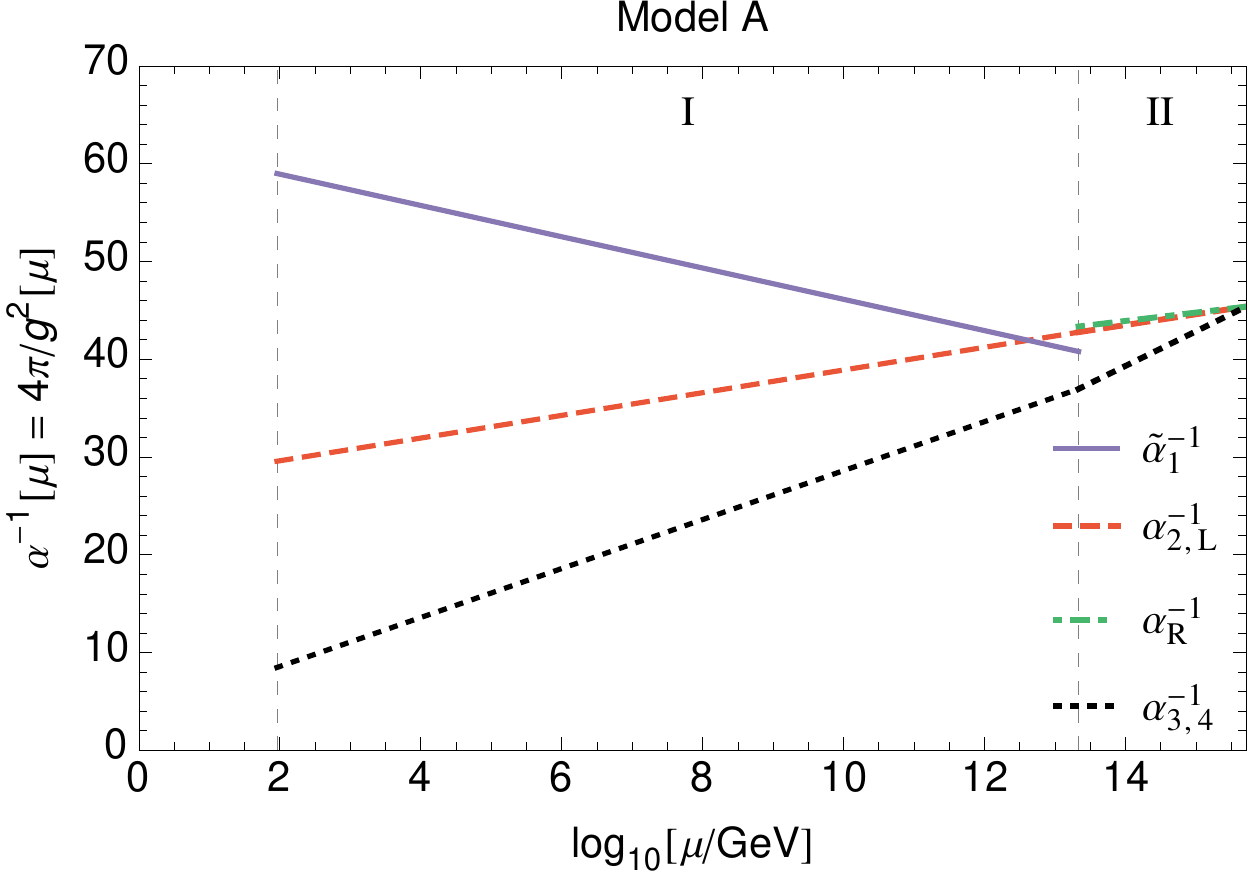}}\hspace{0.4cm}
\subfigure[ ]{\includegraphics[width=6cm]{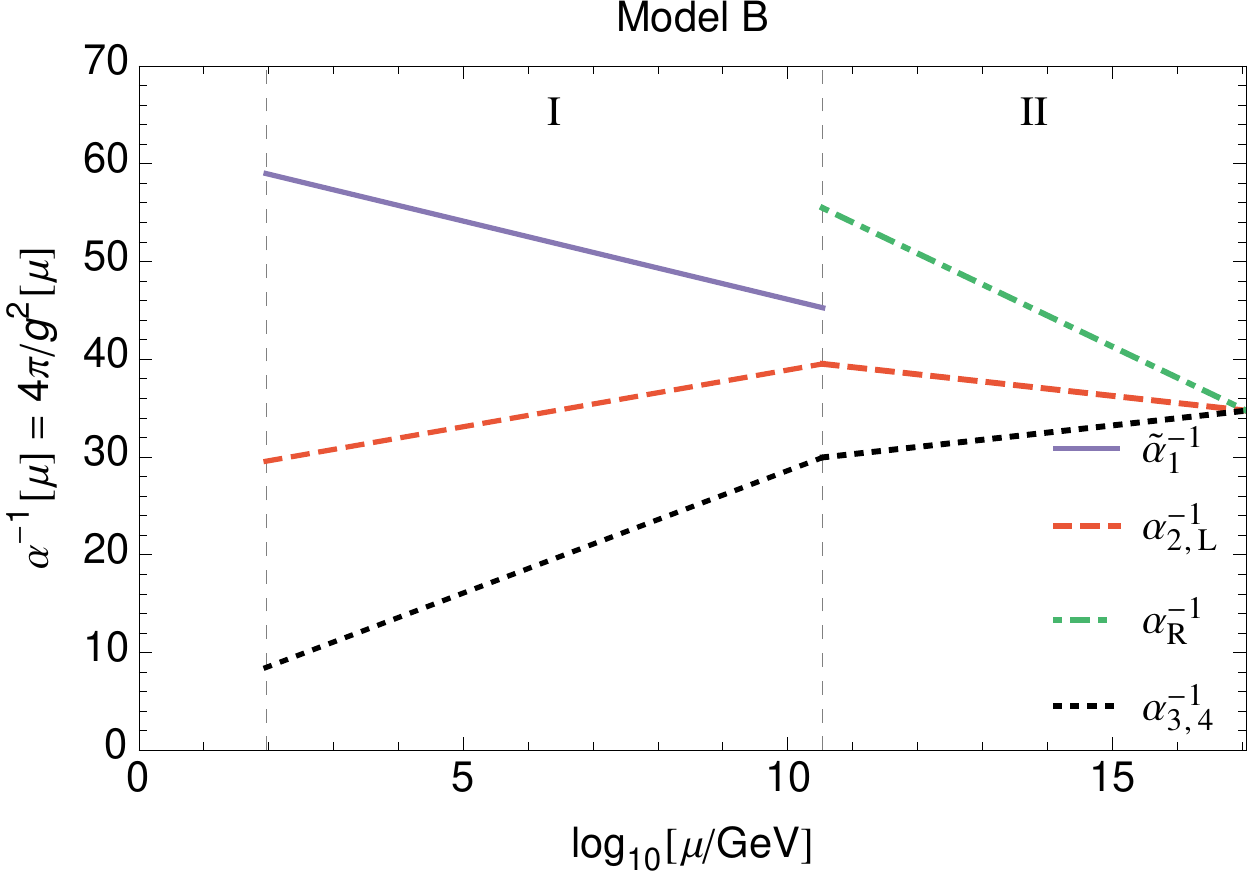}}\vspace{0.3cm}
\subfigure[ ]{\includegraphics[width=6cm]{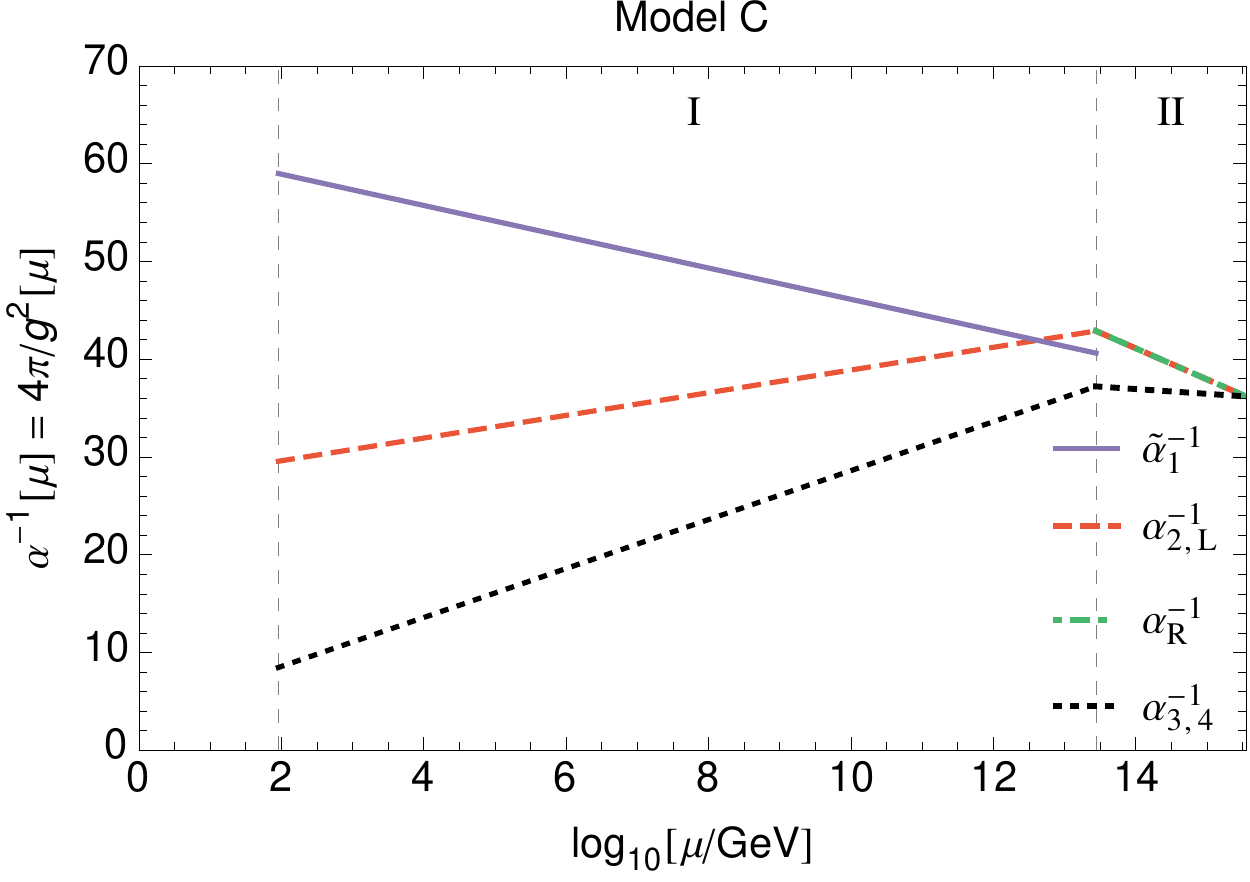}}\hspace{0.4cm}
\caption{Running of the gauge couplings for models A, B, and C. The vertical dotted lines from left to right correspond to the
symmetry breaking scales $M_Z$ and $M_C$,  which also indicate the beginning of the energy intervals I and II, respectively. For $\alpha_1^{-1}$, we plot the redefined quantity $\widetilde{\alpha}^{-1}_1\equiv \dfrac{3}{5}\alpha^{-1}_1$.  Note that in (a), in the interval II, $\alpha^{-1}_L$ and $\alpha^{-1}_R$  evolve very closely but not identically.
}
\label{RGrunning3}
\end{figure}

The particle content and the RG coefficients for the three models
in the two energy intervals are listed in Tables.~\ref{a1A}, \ref{a1B}, and \ref{a1C}.
As stated above, though $\mathcal{S}$ survives in the energy interval I, 
being an SM singlet, it does not contribute to the RG coefficients.
The values of $M_U$, $M_C$, and $\alpha_U$ obtained using the formalism above
are listed in Table.~\ref{Results}.
The running of the gauge couplings for the three models are shown in 
Figure~\ref{RGrunning3}.

We see that for all three models, $M_U$ is below the Planck scale and
$\alpha_U$ is perturbative, as are all the gauge couplings during their course of running.
The value of the symmetry breaking scale $M_C$ is high in the $10^{10\sim 13}\mathrm{GeV}$
range, suggesting that providing $\mathcal{S}$ and $\chi=\Delta_{R3}^{-2/3}\left(1,-4/3,3\right)$ with TeV scale masses, and the TeV scale coupling $\kappa M_R \mathcal{S}\chi^\dagger\chi$ between them would involve fine tuning.

\begin{table}[ht!]
\tbl{The Higgs content and the RG coefficients in the energy intervals for model A.}
{\begin{tabular}{c|l|l}
\hline
$\vphantom{\Big|}$ Interval & Higgs content & RG coefficients
\\
\hline
$\vphantom{\Biggl|}$ II
& $\phi(2,2,1),\;\widetilde{\Delta}_R (1,2,4),\;\Sigma(1,1,15)$
& $\left( a_{L},a_{R},a_{4}\right)^\mathrm{II}
 =\left(-3,\dfrac{-7}{3},\dfrac{-29}{3}\right)$
\\
\hline
$\vphantom{\Biggl|}$   I
& $\phi_2(2,1,1),\; \mathcal{S}(1,0,1),\;\widetilde{\Delta}_{R3}^{-2/3}\left(1,\dfrac{-4}{3},3\right)$
& $\left(a_{1},a_{2},a_{3}\right)^\mathrm{I}=\left(\dfrac{131}{18},\dfrac{-19}{6},\dfrac{-41}{6}\right)$
\\
\hline
\end{tabular}}
\label{a1A}
\end{table}
\begin{table}[ht!]
\tbl{The Higgs content and the RG coefficients in the energy intervals for model B.}
{\begin{tabular}{c|l|l}
\hline
$\vphantom{\Big|}$ Interval & Higgs content & RG coefficients
\\
\hline
$\vphantom{\Biggl|}$ II
& $\phi(2,2,1),\;H(1,1,6),\;\Delta_R (1,3,10),$ 
& $\left( a_{L},a_{R},a_{4}\right)^\mathrm{II}=\left(2,\dfrac{26}{3},-2\right)$
\\
& $\widetilde{\Sigma}(2,2,15)$
& $\vphantom{\Big|}$ 
\\
\hline
$\vphantom{\Biggl|}$   I
& $\phi_2(2,1,1),\; \mathcal{S}(1,0,1),\;\Delta_{R3}^{-2/3}\left(1,\dfrac{-4}{3},3\right)$
& $\left(a_{1},a_{2},a_{3}\right)^\mathrm{I}=\left(\dfrac{131}{18},\dfrac{-19}{6},\dfrac{-41}{6}\right)$
\\
\hline
\end{tabular}}
\label{a1B}
\end{table}
\begin{table}[ht!]
\tbl{The Higgs content and the RG coefficients in the energy intervals for model C.}
{\begin{tabular}{c|l|l}
\hline
$\vphantom{\Big|}$ Interval & Higgs content & RG coefficients
\\
\hline
$\vphantom{\Bigg|}$ II
& $\phi(2,2,1),\;H(1,1,6)\times 2,\;\widetilde{\Sigma}(2,2,15)$ 
& $\left( a_{L},a_{R},a_{4}\right)^\mathrm{II}
 =\left(\dfrac{26}{3},\dfrac{26}{3},\dfrac{4}{3}\right)$ 
\\
& $\Delta_R(1,3,10),\;\Delta_L(3,1,10)$ & \\
& & \\
\hline
$\vphantom{\Biggl|}$   I
& $\phi_2(2,1,1),\; \mathcal{S}(1,0,1),\;\Delta_{R3}^{-2/3}\left(1,\dfrac{-4}{3},3\right)$
& $\left(a_{1},a_{2},a_{3}\right)^\mathrm{I}=\left(\dfrac{131}{18},\dfrac{-19}{6},\dfrac{-41}{6}\right)$
\\
\hline
\end{tabular}}
\label{a1C}
\end{table}

\begin{table}[t!]
\tbl{The predictions of Models A, B, and C.}
{\begin{tabular}{cccc}
\toprule
$\vphantom{\Big|}$
\hspace{1cm}Model \hspace{1cm}
& \hspace{1cm} A \hspace{1cm} 
& \hspace{1cm} B \hspace{1cm} 
& \hspace{1cm} C \hspace{1cm} 
\\
\hline\hline
$\vphantom{\bigg|}$ Unbroken Symmetry
& $G_{224}$ & $G_{224}$ & $G_{224D}$ \\
\hline
$\vphantom{\bigg|}$ $\log_{10}(M_U/\mathrm{GeV})$ &  $15.7$ & $17.1$ & $15.6$ \\
$\vphantom{\bigg|}$ $\log_{10}(M_C/\mathrm{GeV})$ &  $13.3$ & $10.5$ & $13.4$ \\ 
\hline
$\vphantom{\bigg|}$ $\alpha_U^{-1}$ & $45.4$ & $34.7$ & $36.2$ \\
\botrule
\end{tabular}}
\label{Results}
\end{table}

\section{Discussion}

In this letter, we have discussed a possible interpretation of the $750$ GeV diphoton resonance  in the framework of unified $G_{224}$ models derived in the context of a left-right symmetric extension of  non-commutative geometry (NCG) of the Standard Model (SM).
Our framework is a grand unified version of $G_{224}$ models, within which the corresponding Higgs content is restrictively determined (or uniquely determined for each model) from the underlying non-commutative geometry.
This should be contrasted to 
the regular $G_{224}$ models, discussed in the literature, in which the corresponding Higgs context is arbitrarily selected. 

In this note, we have argued that the observed cross sections involving the $750$ GeV diphoton resonance could be realized through a SM singlet scalar field and colored scalars in the NCG of unified $G_{224}$ models. However the color scalars are light and thus fine tuned from the usual effective field theory point of view.
This indicates a certain rigidity of the NCG approach to the Standard Model and its natural completion in the context of the unified $G_{224}$ models. 
As already emphasized, this conclusion is based on the effective field theory reasoning, which might fail in the NCG framewrok due to the
possible mixing between the short-distance and long-distance physics as we have discussed in our previous papers Ref.~\refcite{Aydemir:2015nfa}, as well as Refs.~\refcite{Aydemir:2013zua,Aydemir:2014ama}.
In this paper we have discussed three different scenarios and their implications for the physics beyond the Standard Model. We have concentrated on the purely phenomenological aspects of the NCG unified $G_{224}$ models without relying on
their deep mathematical structure or various novel physics aspects that go beyond the effective field theory framework.

We believe that the discussion presented in this note gives extra evidence to the underlying phenomenological rigidity of the NCG approach towards understanding of the origins of the Standard Model 
and the physics beyond the Standard Model. However, this phenomenological rigidity might be the price one has to pay for the non-commutative nature of the approach, and it might be indicative of a possibly exciting
relation to the non-particle sector of high energy physics, that is to be found in the context of the underlying quantum structure of space and time.

\section*{Acknowledgments}

We thank the Miami winter conference for providing a stimulating environment for the initiation of this project.
The preliminary results of this work have been presented at the meeting \textit{``Noncommutative Geometry, Spectral Action and High Energy Physics,''} in Bruxelles, 27--29 January, 2016, supported by COST Action (QSPACE MP1405). 
We thank the organizers for inviting us to this meeting. In particular, we thank Fedele Lizzi and Walter van Suijlekom for stimulating conversations.
The work of UA is supported by the Swedish Research Council under contract 621-2011-5107 and that of DM is supported in part by the U.S. Department of Energy, grant DE-FG02-13ER41917, task A.

\bibliographystyle{ws-mpla}
\bibliography{NCG}

\end{document}